\documentstyle[prl,aps]{revtex}
\begin{document}
\title{Full Numerical Estimation of Neutrino Mixing Parameters from Solar 
Neutrino Data}
\author{Jai Sam Kim $^{{\rm a}}$\footnote[1]{e-mail: jsk@postech.ac.kr}
and C. W. Kim $^{{\rm b}}$\\
{\it 
$^{{\rm a}}$ Dept of Physics, Pohang University of Science and Technology,\\
Pohang 790-784, Republic of Korea\\
$^{{\rm b}}$ School of Physics, Korea Institute for Advanced Study,\\
Seoul 130-012, Republic of Korea}}
\date{September 17, 1999}
\maketitle

\begin{abstract}

We have numerically computed survival probabilities of solar neutrinos
interacting with matter via MSW mechanism in the full three generation 
formalism with $\Delta m^2_{23}=2.2\times 10^{-3}$ eV$^2$.
For $\theta_{13} \lesssim 30^{\circ}$, we confirm the two regions found in the
two generation case,
(1) with the most likelyhood, small $\sin^2(2\theta_{12}) \sim 0.006$,
(2) with less likelyhood, large $\sin^2(2\theta_{12}) \gtrsim 0.5$.
For $\theta_{13}\gtrsim 30^{\circ}$, we find an additional region,
(3) with even less likelyhood, 
$10^{-5} < \Delta m^2_{12}< 10^{-4}$ eV$^2$ 
and/or $10^{-4} < \sin^2(2\theta_{12}) < 0.5$.
Assuming that the case (1) is valid, we predict that 
$\theta_{13} \simeq 15^\circ$.

\vskip 0.3cm
\noindent
PACS numbers: 14.60.Pq; 26.65.+t; 96.60.Jw
\end{abstract}

\vskip 0.3cm
Recent Super-Kamiokande experiment \cite{SK} shows an evidence for oscillations 
of the atmospheric neutrinos. The data are in good agreement with two-flavor
$\nu_{\mu}\leftrightarrow\nu_{\tau}$ oscillations. These results did establish 
that neutrinos oscillate and possess non-zero masses. Thus full three-flavor 
neutrino oscillation analysis is needed for more accurate determination of 
neutrino masses and mixing anlges. There are many investigations of neutrino 
oscillations with three neutrino generation schemes of solar neutrinos, 
accelerator and reactor neutrinos as well as atmospheric neutrinos 
\cite{TH1,TH4,TH5}. 

The solar neutrino puzzle is the discrepancy between the standard solar model 
predictions \cite{BP95,BBP98} and the results of the solar neutrino experiments.
The standard solar model of Bahcall and Pinsonneault predicts that the neutrino 
event rates are $9.3^{+1.2}_{-1.4}$ SNU for the chlorine experiment, 
$137^{+8}_{-7}$ SNU for the gallium experiment, and
$(6.62^{+0.93}_{-1.12})\times 10^6$ /cm$^2$/sec for the Super-Kamiokande 
experiment.
The observed neutrino event rates are
$2.56 \pm 0.16 \pm 0.15$ SNU for Homestake Chlorine experiment \cite{HOM},
$77.5 \pm 6.2^{+4.3}_{-4.7}$ SNU for GALLEX Gallium experiment \cite{GAL},
$67.2^{+7.2+3.5}_{-7.0-3.0}$ SNU for SAGE Gallium experiment \cite{SAG}, and 
$(2.44 \pm 0.05^{+0.09}_{-0.07})\times 10^6$ /cm$^2$/sec for the Kamiokande 
experiment \cite{KAM}.

Several mechanisms were proposed for the depletion of the solar neutrinos.
Among them the Mikheyev-Smirnov-Wolfenstein (MSW) effect \cite{WOL,MIK} has 
been most popular because it induces large flavor transitions in spite of small 
vacuum mixing angle in an energy dependent way. Two-flavor MSW solutions 
\cite{BKS,HATA} to the solar neutrino puzzle was fully analysed with high 
accuracy, including both the day/night effect and spectral information.

In the two-neutrino scheme with the MSW effect it was found that 
there are two regions in the $\sin^22\theta-\Delta m^2$ parameter space 
that yield two different solutions of the solar neutrino problem:
the small mixing angle solution and the large mixing angle solution.
The chi-square analysis indicates that the small mixing angle
solution is favored \cite{BKS}.
Authors in ref. \cite{GKLN} showed that the high-$\Delta m^2$ part
of the large mixing angle MSW solution of the solar neutrino problem
is disfavored by the Super-Kamiokande atmospheric neutrino data
assuming the scheme of neutrino mixing indicated by the result of 
the reactor neutrino oscillation experiment CHOOZ \cite{CHOOZ}. 
This implies that the mixing matrix element $U_{e3}$ (or $\theta_{13}$ in 
our parametrization which is the same as that of the Particle Data Group 
\cite{pdg}) is small and the oscillations of solar and atmospheric neutrinos 
are decoupled under the assumption of neutrino mass hierarchy 
$\Delta m^2_{12} \ll \Delta m^2_{23}$,
($\Delta m^2_{12}\equiv m^2_2 - m^2_1, \Delta m^2_{23}\equiv m^2_3-m^2_2$).

Since the solar neutrino oscillates into two other flavors, a full analysis 
may result in some new findings. Especially the angle $\theta_{13}$ may play 
a non-trivial role. There are many papers in which the MSW effect is considered
in three-generation-neutrino scheme \cite{par,kuo,cwk,bal,msm,bahax,FOG,BAR}. 
However, they used the Landau-Zenor approximate formula for treating flavor 
conversion in the non-adiabatic resonant region of the solar interior with 
the assumption that all neutrinos are created at the center of the Sun.
Two groups \cite{HAX,FIOR} implemented numerical methods with two neutrino
flavors.

This method of using the Landau-Zenor formula is unreliable when the resonant 
region and the neutrino production region overlap. 
In order to demonstrate the situation, we first present a sample of
our numerical analysis which shows that for low energy neutrinos the survival 
probabilities indeed depend very much on their creation sites, $r$. 
The survival probability of a solar neutrino with energies, $0.3<E_\nu<1.0$ 
MeV, varies rapidly in the range, $0<r/R_\odot<0.2$. 
Thus $r$-dependence of the survival probability should be taken into account
for the $pp$ and Be neutrinos though not as much for the boron neutrinos. 
That is, one should consider the radial distribution of the solar neutrinos 
in analysing the gallium experiment data \cite{GAL,SAG,GNO} and the Borexino 
data \cite{BOR}.

Recently we \cite{KIM} presented a numerical algorithm for computing the 
survival probability of an electron neutrino in its flight through the solar 
core experiencing the MSW effect. We adopted a hybrid method to integrate the 
neutrino evolution equation in the full three-generation formalism. 
In order to reduce the number of integrations we used the importance sampling 
method for sampling the neutrino creation energy and position. In order to
reduce the amount of computation in each integration we need to find the 
optimum radii at which numerical integration is started and stopped. 
Outside these radii we would use the adiabatic conversion formula. 
We checked the location of the non-adiabatic resonant regions, where direct
numerical integration is performed, by checking if the relative ratios of
the off-diagonal and diagonal elements of the Hamiltonian are below a
certain limit, say $\gamma_c$. Thus the frequency and amount of numerical 
integrations are greatly reduced. Even with these tactics it is 
extremely time consuming to solve the neutrino evolution equations for the 
three-generation case in a large region of the parameter space and 
it requires deployment of a parallel supercomputer. 
We developed a parallel algorithm for a message passing parallel computer.

In this paper, we present results of our lengthy numerical computation of
the solar neutrino survival probabilities.
We will show that the numerical MSW solutions to the solar neutrino problem
within the three-generation formalism lead to a new constraint to
the neutrino masses and mixing angles.
In the three neutrino formalism, we have three vacuum mixing angles 
($\theta_{12},\theta_{23},\theta_{13}$) and one phase to describe mixing
between mass and flavor eigenstates.
The angle $\theta_{23}$ is irrelevant in the solar neutrino problem and the 
CP violating phase may be ignored. Thus adjustable parameters are two mixing 
angles and two mass squared differences.
In conformity with the Super-Kamiokande data on the atmospheric neutrinos, 
we fixed two parameters, 
$\Delta m^2_{23}=2.2\times 10^{-3}$ eV$^2$ and $\theta_{23}=43.5^{\circ}$
and searched a wide range of the parameter space spanned by
$\Delta m^2_{12}$, $\theta_{12}$, and $\theta_{13}$. 
We used a 24$\times$21 logarithmic grid for $\Delta m^2_{12}$ and 
$\sin^2(2\theta_{12})$.
We focused on the interesting region,
$2\times 10^{-6} \leq \Delta m^2_{12} \leq 4\times 10^{-4}$ eV$^2$ and
$10^{-4} \leq \sin^2(2\theta_{12}) \leq 1$.

In our production run, we refined the importance sampling method for sampling
neutrino creation energies and radii. For the boron and CNO neutrinos,
we chose 64 energy sample points ranging from 0.23 MeV to 18.65 MeV and 
60 radius sample points ranging from 0 to 0.353$R_\odot$.
The three energy values, 0.384, 0.862, 1.442 MeV, are included to evaluate
the Be and pep neutrinos more accurately.
For the most numerous and model independent $pp$ neutrinos, to be counted by 
the gallium detectors, we chose 60 energy sample points ranging from 0.23 MeV 
to 0.4293 MeV and 36 radius sample points ranging from 0 to 0.4$R_\odot$.
In order to determine the radial limits of the non-adiabatic region,
we used a very generous critical value $\gamma_c=0.005$, which guaranteed the 
accuracy of 0.001 for the survival probabilities in the worst case.

We did not include the Earth effect in the calculation. The sensitive parameter 
region to the Earth density is around $\Delta m^2_{12}\sim 10^{-6}$ eV$^2$.
In the two-generation scheme the Earth effect does not cause much change of the 
combined parameter regions for the solar neutrino problems.
The data of energy spectrum and day/night variation from the Kamiokande and 
Super-Kamiokande may give more restriction on the parameter space of neutrino 
mass and mixing angle. The absence of any day/night variation in the data is 
consistent with neglecting the MSW resonance effect in the Earth for the 
electron neutrino. The excluded region by the data of neutrino energy spectrum 
is around $\Delta m^2_{12}\sim 10^{-4}$ eV$^2$ which is outside the combined 
parameter regions in the two-generation case.

To compute the event rates, we used the 98 SSM flux data taken from 
\cite{BP95,BBP98}.
We present our results for several values of $\theta_{13}$ in the familiar 
iso-SNU/FLUX contour format in Figs. 1.
In these figures only the mean SNU(FLUX) values are plotted.
From these figures we notice three important facts: 
(1) When $\theta_{13} \lesssim 22.5^\circ$, the small mixing angle region
where $\sin^2(2\theta_{12})\ll 1$, is the only region where all three curves 
can come close. In the large mixing angle regions two curves can cross
at a time but the crossing points are rather far apart.
(2) Beyond $\theta_{13} \gtrsim 22.5^\circ$, the gap which the Ga curve passes 
between the Cl curve and the Water curve becomes wide and even the small mixing 
angle solution becomes less likely.
(3) When $\theta_{13} \gtrsim 35^\circ$, there is no place in the considered 
region of the $\Delta m^2_{12} - \sin^2(2\theta_{12})$ plane where any pair 
of curves can meet. Thus the likelyhood to have a solution with a large (1-3)
mixing angle is very small.

As we can see in the figures the chlorine data are least sensitive to 
$\theta_{13}$. The Water data are a little more sensitive to $\theta_{13}$ 
but the shape of the curve does not change. In the small $\Delta m^2_{12}$ 
region with small values of $\theta_{13}$ the Water curves overlap the Cl 
curves as both experiments are sensitive mostly to the boron neutrino. 
Thus it is impossible to determine the mixing angles decisively from these 
two experiments. However, the gallium detector is sensitive to the $pp$ 
neutrinos also and its iso-SNU curve crosses the other two curves at large 
angles and thus most useful for a precision determination of $\theta_{12}$.
The figures show that the gallium curves are also very sensitive to a choice of 
$\theta_{13}$ and they can be used to impose a non-trivial constraint on 
$\theta_{13}$. 

Experimental data contain two kinds of errors, statistical and systematic. 
We take weighted averages and estimate combined errors as reviewed in 
\cite{pdg}.
For the Cl experiment, we use $R_{{\rm Cl}} = 2.56\pm0.23$ SNU and 
for the Water experiment, we used
$R_{{\rm Water}} = (2.44\pm0.10)\times 10^6$/cm$^2$/sec. 
For the two gallium experiments we take weighted average to get 
$R_{{\rm Ga}} = 73.1\pm6.6$ SNU.
The flux data from the SSM model contain some theoretical uncertainties.
Thus we need to represent the 1$\sigma$ deviation values of SNU(FLUX). 
Figs. 2 take into account both theoretical uncertainties and experimental
errors. 
The upper curves are from the lower limit of the SSM flux data and the upper 
values of detection rates of each experiment and vice versa.

For small values of $\theta_{13}$, our result is consistent with the two 
generation case \cite{BKS,HATA} as expected. Barbieri {\it et. al.} \cite{BAR} 
performed extensive analysis fo solar neutrino experiments with all three
neutrinos. They used the Landau-Zenor approximation formula given in 
\cite{par,kuo}. The experimental data are slightly updated since their work. 
They used, $R_{{\rm Cl}} = 2.54\pm0.20$ SNU for Cl, 
$R_{{\rm Ga}} = 75\pm 7$ SNU for Ga, and
$R_{{\rm Water}} = (2.51\pm0.16)\times 10^6$/cm$^2$/sec for Water.
So direct comparison of our results with their results is not possible.
However, we made iso-SNU(FLUX) plots with these data also. 
For $\theta_{13}=1.5^\circ,\ 15^\circ$, our best estimates of $\Delta m^2_{12}$
in units of $10^{-6}$ eV$^2$ and  $\sin^2(2\theta_{12})$ in units of $10^{-3}$
are $(5.05,\ 5.89)$ and $(6.50,\ 3.99)$ respectively. 
However, for $\theta_{13}=30^\circ$ our estimates are $(7.44,\ 2.20)$,
whereas their plots indicate $(\sim 8,\ \sim 6)$.
It seems that their estimates of $\Delta m^2_{12}$ are slightly larger than 
ours but still within the 1$\sigma$ limit. For small values of $\theta_{13}$ 
we agree with theirs within the 1$\sigma$ limit whereas for large
values our estimate of $\sin^2(2\theta_{12})$ is much smaller. 
In both cases, for large $\theta_{13}$, there is a 1$\sigma$ region where
$\Delta m^2_{12}\sim 10^{-4}\ {\rm eV}^2$.

Due to the theoretical uncertainties and experimental errors, all angular 
values of $\theta_{13}$ can accomodate all three experiments in broad ranges
of $\Delta m^2_{12}$ and $\theta_{12}$. For small $\theta_{13}$
the `small mixing angle' solution gives the most likely solution, with a
small patch of `large mixing angle' region hanging in the upper-right corner.
However, for large $\theta_{13}$, the large boomerang shaped region in Figs. 2 
allows broad range of values for $\Delta m^2_{12}$ and $\sin^2(2\theta_{12})$.
Large values of $\Delta m^2_{12}\approx 10^{-4}$ eV$^2$ are possible 
and $2\cdot 10^{-4} < \sin^2(2\theta_{12})<0.2$ for $\theta_{13}=40^\circ$.
But its likelyhood seems to be minimal as Figs. 1 indicate.
However, the earth effect can broaden the large $\theta_{12}$ region, 
which we did not consider in this work.

We repeated the computation with $\Delta m^2_{23}=5.0\times 10^{-4}$ eV$^2$,
which is the $1\sigma$ lower limit of the Super-Kamiokande estimates.
For small values of $\theta_{13} < 22.5^\circ$ the contours of mean SNU values 
(Fig. 1) look almost the same as in the case 
$\Delta m^2_{23}=2.2\times 10^{-3}$ eV$^2$. 
However, in the latter case ($\Delta m^2_{23}=5.0\times 10^{-4}$ eV$^2$),
for large values of $\theta_{13}$ the Water contours are shifted slightly to 
the left and the Ga contour are pushed further to the upper-left direction, 
while the Cl contours are intact. Nevertheless we do not see noticable 
differences between the two cases.
However, with $\Delta m^2_{23}=2.0\times 10^{-4}$ eV$^2$, we noticed 
non-trivial deviations. In this case (2-3) transitions can take place in 
addition to the (1-2) transition.
Our program has been improved to compute the survival probabilities even for
the most general situation when all three transitions occur.
We repeated the computation with the same set of $\theta_{13}$.
The most significant difference is that the $1\sigma$ region broadens for 
large $\theta_{13}$.

We give the values of the small (1-2) mixing angle solutions within 1$\sigma$
range in Table 1. It is interesting to notice that $\chi^2_{{\rm min}}$ 
depends on $\theta_{13}$. Its minimum value occurs when the Cl and the Water 
curves overlap perfectly. It can be used to predict the most likely value 
of $\theta_{13}$ from solar neutrino data alone. In this case the minimum 
occurs at $\theta_{13}=13.21^\circ$.

\hskip 0.3cm
\begin{center}
\begin{quote}
Table 1. Best estimates of small (1-2) mixing parameters for several values of
$\theta_{13}$ obtained by minimizing $\chi^2$. Here, we used the definition
$\chi^2 = \sum_i((R_i-R_i^{{\rm exp}})/\Delta R_i^{{\rm exp}})^2$
with $i={\rm Cl,\ Ga,\ Water}$, which is caused purely by the experiments.
The 1$\sigma$ bounds are also indicated.
\end{quote}
\vskip 0.3cm
\begin{tabular}{|c|c|c|c|}
\hline
$\theta_{13}$ (in deg) & $\Delta_{12}\times 10^6$ eV$^2$ & $\sin^2(2\theta_{12})\times 10^3$ & $\chi^2_{{\rm min}}$\\
\hline
1.5 & $ 4.81^{+2.62}_{-0.80}$ & 6.29$^{+2.57}_{-2.69}$ & 0.04526 \\
5   & $ 4.89^{+2.79}_{-0.78}$ & 6.10$^{+2.56}_{-2.76}$ & 0.05902 \\
10  & $ 5.52^{+2.78}_{-1.42}$ & 5.22$^{+2.92}_{-2.44}$ & 0.07068 \\
12.5& $ 6.19^{+2.15}_{-2.18}$ & 4.50$^{+2.12}_{-3.15}$ & 0.00237 \\
15  & $ 6.62^{+2.44}_{-2.35}$ & 4.00$^{+3.17}_{-1.65}$ & 0.03779 \\
22.5& $ 7.59^{+2.90}_{-1.94}$ & 2.81$^{+1.49}_{-1.06}$ & 0.76387 \\
30  & $ 7.58^{+8.29}_{-0.63}$ & 2.12$^{+1.12}_{-1.01}$ & 3.36182 \\
\hline
\end{tabular}
\end{center}
\vskip 0.3cm
In summary we have computed the solar neutrino survival probabilities
$P(\nu_e\rightarrow\nu_\beta; E_\nu,r; \Delta m^2_{ij},\theta_{ij})$ 
(1) within the full three generation framework
(2) using accurate numerical methods and 
(3) included the effects due to neutrino creation positions to full extent.
We have found that there are three regions in the parameter space where
all three solar neutrino experiments can be accomodated within the 1$\sigma$ 
limit:
for small (1-3) mixing angles $\theta_{13} \lesssim 30^{\circ}$,
(1) with the most likelyhood,
small $\Delta m^2_{12}\sim 5\times 10^{-6}$ eV$^2$ and
small $\sin^2(2\theta_{12}) \sim 0.006$,
(2) with less likelyhood,
large $\Delta m^2_{12}\sim 10^{-4}$ eV$^2$ and
large $\sin^2(2\theta_{12}) \gtrsim 0.5$;
and for large (1-3) mixing angles $\theta_{13}\gtrsim 30^{\circ}$,
(3) with even less likelyhood,
$10^{-5} < \Delta m^2_{12}< 10^{-4}$ eV$^2$
and/or $10^{-4} < \sin^2(2\theta_{12}) < 0.5$.

From Figs. 1 and Figs. 2 and Table 1, we conclude that the small 
$\sin^2(2\theta_{12})$ solution with $\theta_{13}$ smaller than $22.5^\circ$ 
will cope with future experiments more likely than the large 
$\sin^2(2\theta_{12})$ solutions. 
We estimate the most likely value of $\theta_{13}$ by minimizing 
$\chi^2_{{\rm min}}$ with respect to $\theta_{13}$, which comes out to be 
$\theta_{13}=13.21^\circ$ from the Table 1.
Our conclusion is consistent with the CHOOZ estimate of $\theta_{13}<15^\circ$.
If the errors in the Cl and the Water experiments are reduced further
and the value of $\theta_{13}$ measured in CHOOZ and other experiments agree 
with our prediction, it can serve as an interesting test of the Standard Solar 
Model and the MSW mechanism.

We would like to thank Dr. J.D. Kim for discussions and Ms. K.S. Lee
for making plots and analysing data for the table. We also thank Dr. S.B. Kim 
for clarifying the Super-Kamiokande data.
This work was funded by the POSTECH research fund.

\end{document}